\documentclass[12pt,showpacs,preprintnumbers,amsmath,amssymb,superscriptaddress,floatfix]{iopart}
\usepackage{graphicx}% Include figure files
\usepackage{dcolumn}% Align table columns on decimal point
\usepackage{bm}% bold math
\usepackage{longtable}
\usepackage{array}
\usepackage[usenames]{color}
\usepackage{caption}

\begin{document}
 \title{A Cross-correlation method to search for gravitational wave bursts with AURIGA and Virgo}
\bigskip
\address{\textbf{THE AURIGA COLLABORATION}}
\author{
 M. Bignotto$^{1,2}$, M. Bonaldi$^{3,4}$, M. Camarda$^{5}$ ,
 M. Cerdonio$^{1,2}$,  L. Conti$^{1,2}$,  M. Drago$^{1,2}$,
 P. Falferi$^{3,4}$,  N. Liguori$^{1,2}$, S. Longo$^{6}$,
  R. Mezzena$^{4,7}$,  A. Mion$^{4,7}$,
  A. Ortolan$^{6}$, 
  G.A. Prodi$^{ 4,7}$, V. Re$^{4,7}$,
 F. Salemi$^{4,7}$,  L. Taffarello$^{2}$, G. Vedovato$^{2}$, A. Vinante$^{3,4}$, S. Vitale$^{4,7}$, J.P. Zendri$^{2}$ }

\address{$^1$ Dipartimento di Fisica, Universit\`a di Padova, Via Marzolo 8, 35131 Padova, Italy}
\address{$^2$ INFN, Sezione di Padova, Via Marzolo 8, 35131 Padova, Italy}
\address{$^3$ Istituto di Fotonica e Nanotecnologie, CNR-Fondazione Bruno Kessler, I-38050 Povo (Trento), Italy}
\address{$^4$ INFN, Gruppo Collegato di Trento, Sezione di Padova, I-38050 Povo, Trento, Italy}
\address{$^{5}$ Dipartimento di Ingegneria Informatica, Universit\`a di Padova, Via G. Gradenigo 6a, 35131 Padova, Italy}
%\address{$^{18}$ Dipartimento di Fisica, Universit\`a di Ferrara and INFN, Sezione di Ferrara, I-44100 Ferrara, Italy}
\address{$^{6}$ INFN, Laboratori Nazionali di Legnaro, 35020 Legnaro, Padova, Italy}
\address{$^{7}$ Dipartimento di Fisica, Universit\`a di Trento, I-38050 Povo, Trento, Italy}
%\address{$^{26}$ Consorzio Criospazio Ricerche, I-38050 Povo, Trento, Italy}
\bigskip
\address{\textbf{THE Virgo COLLABORATION}}
\author{F.Acernese $^{7,9}$,
M.Alshourbagy$^{15,16}$,
P.Amico$^{13,14}$,
F.Antonucci$^{19}$,
S.Aoudia$^{10}$,
P.Astone$^{19}$,
S.Avino$^{7,8}$,
L.Baggio$^1$,
G.Ballardin$^2$,
F.Barone$^{7,9}$
% ...
}

\address{$^1$Laboratoire d'Annecy-le-Vieux de Physique des Particules (LAPP),  IN2P3/CNRS, Universit\'e de Savoie, Annecy-le-Vieux, France}
\address{$^2$European Gravitational Observatory (EGO), Cascina (Pi), Italia.}
\address{$^3$INFN, Sezione di Firenze, Sesto Fiorentino, Italia.}
\address{$^4$ Universit\`a degli Studi di Firenze, Firenze, Italia.}
\address{$^5$ Universit\`a degli Studi di Urbino "Carlo Bo", Urbino, Italia.}
\address{$^6$LMA, Villeurbanne, Lyon, France.}
\address{$^7$ INFN, sezione di Napoli, Italia.}
\address{$^8$ Universit\`a di Napoli "Federico II" Complesso Universitario di Monte S.Angelo, Italia.}
\address{$^9$ Universit\`a di Salerno, Fisciano (Sa), Italia.}
\address{$^{10}$Departement Artemis -- Observatoire de la C\^ote d'Azur, BP 4229 06304 Nice, Cedex 4, France.}
% ...
\address{$^{13}$INFN, Sezione di Perugia, Italia.}
\address{$^{14}$Universit\`a di Perugia, Perugia, Italia.}
\address{$^{15}$INFN, Sezione di Pisa, Italia.}
\address{$^{16}$ Universit\`a di Pisa, Pisa, Italia.}
% ...
\address{$^{19}$INFN, Sezione di Roma, Italia.}
% ...

\bigskip 
\address{e-mail virginia.re@lnl.infn.it}

\date{\today}% It is always \today, today,
             %  but any date may be explicitly specified

\begin{abstract}

We present a method to search for transient GWs using a network of detectors with different spectral and directional sensitivities: the interferometer Virgo and the bar detector AURIGA. The data analysis method is based on the measurements of the correlated energy in the network by means of a weighted cross-correlation. To limit the computational load, this coherent analysis step is performed around time-frequency coincident triggers selected by an excess power event trigger generator tuned at low thresholds. The final selection of GW candidates is performed by a combined cut on the correlated energy and 
on the significance as measured by the event trigger generator. The method has been tested on one day of data of AURIGA and Virgo during September 2005. The outcomes are compared to the results of a stand-alone time-frequency coincidence search. We discuss the advantages and the limits of this approach, in view of a possible future joint search between AURIGA and one interferometric detector.

\end{abstract}
%\maketitle
\section{Introduction}
\label{intro}
We present a study on the performances of a gravitational wave (GW) observatory composed by a hybrid network of detectors. In particular, we focus on %This study finds its scientific motivations in 
the possibility to use a resonant detector to perform GW observations with one interferometric detector. This could be an opportunity in the scenario after LIGO S5 run and the first Virgo science run, when most of the interferometers will be shut down for upgrading: current plans are that GEO will be kept in operation till the start of the LIGO S6 and the second Virgo science runs, supported by LIGO Hanford 2k detector over weekends. In this sense, we present a case study  on joint observations between AURIGA and Virgo on a test period of 24 hrs. 

In the past years, various searches for GW signals have been independently performed by networks of resonant bars \cite{IGECPRL,IGEC,IGEC2005} or interferometers \cite{S2,S4}.
There have been also some attempts to perform burst searches among detectors with  different spectral sensitivity and orientation: by TAMA and the LIGO Scientific Collaboration (LSC) \cite{LIGOTAMA}, by AURIGA and the LSC \cite{AURLIGOgwdaw,AURLIGOamaldi6,AURLIGOfinal} and by the INFN bars and the Virgo Collaboration \cite{virgobars}.

The proposed network search strategy takes as a starting point the WaveBurst+CorrPower \cite{rstat,CorrPower} search used by LSC for the S3 and S4 analyses \cite{S3,S4}. That search was greatly innovative: a two-step search composed of an ExcessPower-like event trigger generator plus a cross-correlation test which allowed an efficient reduction of false alarms. In that case, however, the detectors partecipating to the network were almost aligned and had a similar spectral sensitivity. 
An extension of such methodology to the case of similar but misaligned detectors has been discussed in literature \cite{Klimenko}. The novelty of our work consists in a further generalization to detectors with different spectral sensitivities, so that it can be implemented between a resonant bar and an interferometer. 
To better characterize the method, we compare its performances with those of a simple time-frequency coincidence search.

The paper is organized as follows: in section 2 we introduce the search method. Section 3 presents an overview of the
exchanged data and summarizes the main steps of the network pipeline and of the tuning performed on chosen test-statistics. Results and conclusions are presented in section 4 and 5 respectively.

\section{The search method}

The GW search method described in this paper is characterized by two main parts: the event trigger generator, whose role is to select a subset of {\it interesting} triggers and a coherent analysis. The trigger search is
based on Waveburst \cite{Waveburst}, an excess power algorithm based on the wavelet decomposition in the time-frequency plane. In the present work, Waveburst has been used in coincidence mode, i.e. the algorithm selects time-frequency coincident excesses of power between the two detectors. 
The step of coherent analysis is based on a cross-correlation test between data streams weighted by a combination of the strain sensitivities of the two detectors (XCorr).

\label{method}

Our method assumes that the GW components at earth can be parametrized as 
\begin{equation}\label{h_dec}
h_+ (t) = h_0 (t) \cdot \cos[\psi(t)]~~~~~~~h_\times (t) = \epsilon \cdot h_0 (t) \cdot \sin[\psi(t)]
\end{equation}
where $h_0(t)$ and $\psi(t)$ are the time-varying amplitude and time-varying phase, common to both polarization components, and $\epsilon$ is the ratio of the cross and plus amplitudes. A large class of GW signals can be parametrized as described above, including linearly, elliptically and circularly polarized GWs, even with sweeping frequencies. 

\begin{figure}[!h]
 \begin{center}
  \includegraphics[width=120mm]{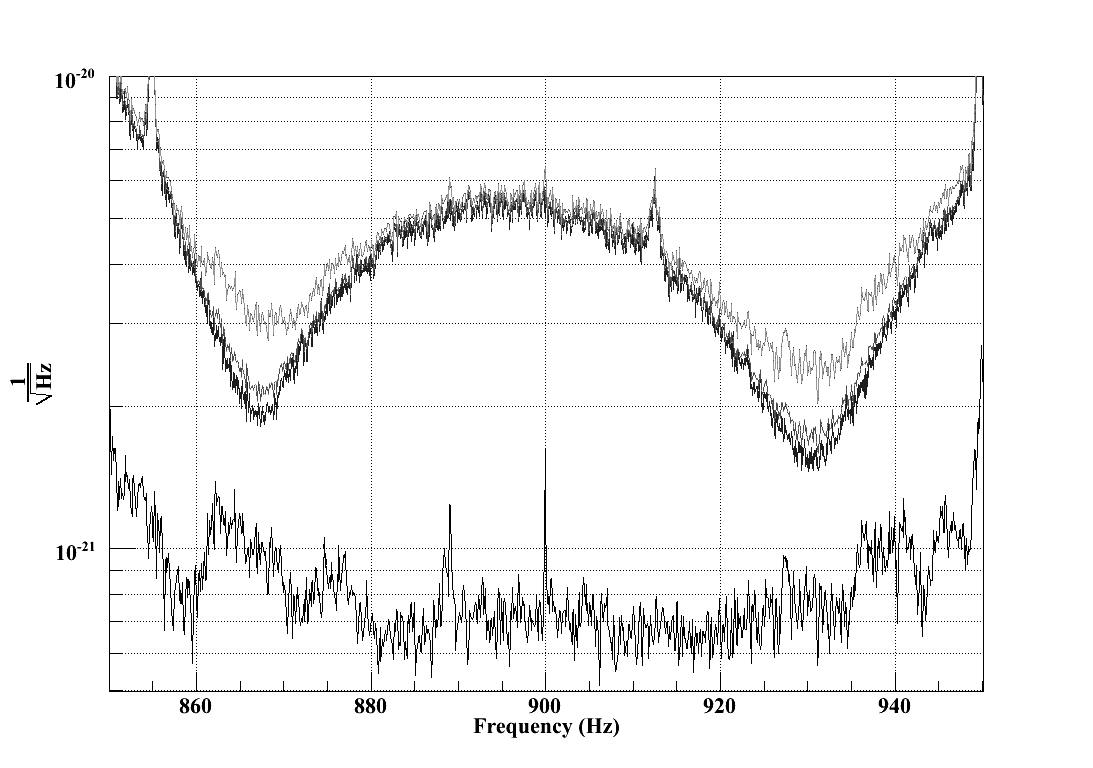}
 \end{center}
\caption{\small{ Strain noise power spectral densities of AURIGA and Virgo detectors on September the $15^{th}$, 2005 (black continuous lines). The gray curve shows the filter $\sigma_{null}(\omega)$ with the approximation $R_{AU} = R_V$. The exact shape of the spectral filter is lower limited by the AURIGA strain noise, corresponding to sources with $R_{AU} \ll R_V$. The light gray curve shows the exact spectral filter for sources with $R_{AU} = 2.4 R_V$ and can be interpreted as an upper limiting curve since the probability that $R_{AU} \leq 2.4 R_V$ is 90\% for a source distributed in the Milky Way with random polarization. We conclude that the spectral filter with $R_{AU} = R_V$ is a satisfactory approximation for our purposes.
}}\label{xcorr-filter}

\end{figure}

The strain produced on the detector $\alpha$ by an incoming burst signal with polarization components in the wavefront frame $h_{+,\times}(t)$ is:
\begin{equation}\label{Gen_h}
h_\alpha (t) = F_{\alpha +} \cdot h_+(t) + F_{\alpha \times} \cdot h_\times(t)
\end{equation}
where $F_{\alpha +}$ and $F_{\alpha \times}$ are the \textit{antenna pattern} functions \cite{JKS} (and references therein).
Following \cite{tinto}, eq.\ref{Gen_h} becomes:
\begin{equation}\label{hR}
h_\alpha (t) = h_0 (t) \cdot R_\alpha (\theta, \phi, \epsilon)\cdot \cos[\psi(t)+\xi_\alpha(\theta, \phi, \epsilon)]
\end{equation}
where $R_\alpha$ is a directional sensitivity, $\xi_\alpha$ is a phase shift and ($\theta, \phi$) is the location of the GW source \footnote{With some algebra, it is easy to find that $R_\alpha = \sqrt{(F_{\alpha +})^2 + (F_{\alpha \times}\cdot \epsilon)^2}$ and $\xi_\alpha= -arctan[\frac{F_{\alpha \times}\cdot \epsilon}{F_{\alpha +}}]$.}. 
The reconstructed strain at the input of two detectors, $\alpha$ and $\beta$, is $x_{\alpha,\beta}(t)=h_{\alpha,\beta}(t)+ n_{\alpha,\beta}(t)$, where $n_{\alpha,\beta}$ are the two independent noises. It has been shown in \cite{chatterji-2006-74} that the following linear combination of the two reconstructed strains, called \textit{null stream}, cancels the signal: $x_{null}(t)\equiv x_\alpha(t) R_\beta -  x_\beta(t+t') R_\alpha$, where $t'$ includes the light travel time and a suitable fraction of the typical GW period, so that $\psi(t)+\xi_{\alpha}=\psi(t+t')+\xi_{\beta}$\footnote{An additional assumption is required here: the envelope of the GW signal be smooth in time so that $h_0(t) \simeq h_0(t+t')$. Whenever one considers a cross-correlation with a narrow-band detector, this approximation is automatically verified since the reconstructed strain at input has to be bandlimited by a suitable filter (see fig.\ref{xcorr-filter}).}. %In case the two detectors are at different sites, the only difference is that $t'$ has to include also the light travel time.

We introduce the Fourier transform of the null stream and normalize it to its variance per unit bandwidth,  $\sigma_{null}^2(\omega)$:  we properly filter $x_{\alpha}(t)$ obtaining the \textit{weighted} reconstructed strain at detector $\alpha$:
\begin{equation}
\label{cross-freq}
\widetilde{x}_{\alpha, w}(\omega)= \frac{\widetilde{x}_\alpha(\omega) }{\sigma_{null}(\omega)} = \frac{\widetilde{x}_\alpha(\omega) }{\sqrt{S_\alpha(\omega) R_\beta^2 + S_\beta(\omega) R_\alpha^2}} 
\end{equation}
 where $S_{\alpha, \beta}$ are the noise power spectral densities of the detectors in terms of GW strain.
Hence, the normalized null stream is $x_{null, w}(t)\equiv x_{\alpha,w}(t) R_\beta -  x_{\beta,w}(t+t') R_\alpha$.

One well known method to search for GW signals in the data relies on the minimization of the null energy \cite{tinto, chatterji-2006-74}, $E_{null}\equiv \int dt x_{null,w}^2(t)$, where the time integral is performed on the signal duration plus any typical response time of the narrower band detector. The null energy can be expressed in terms of the correlated energy and the incoherent energy of the network: $E_{null} = - E_{cor} + E_{inc}$. The former is the contribution of the cross-correlation of the detectors, $E_{cor}= 2 \int dt  x_{\alpha,w}(t) x_{\beta,w}(t+t') R_\alpha R_\beta$. The latter is the auto-correlation contribution of the detectors, $E_{inc}= \int dt (x_\alpha^2(t) R_\beta^2 + x_\beta^2(t+t') R_\alpha^2 )$. As discussed in \cite{chatterji-2006-74}, a GW candidate is selected against background events more efficiently by maximizing $E_{cor}$ rather than by minimizing $E_{null}$. In fact, $E_{null}$ can take low values even for accidental events with small $E_{cor}$ and $E_{inc}$; instead, for detectable GW signals, we expect a higher $E_{inc}$, almost balanced by a positive $E_{cor}$. 
For these reasons, this coherent step of network analysis is based on the maximization of the correlated energy $E_{cor}$ in our null stream. 

In principle, $E_{cor}$ depends on $\theta, \phi$ and $\epsilon$ of the source through $t'$, $R_\alpha$ and $R_\beta$. However, we checked that in the case of random polarized GW radiation, emitted by sources distributed either in the galaxy or uniformly in the sky, we can follow an approximated maximization procedure of $E_{cor}$ assuming $R_\alpha \simeq R_\beta$. 

The main reason is that AURIGA is limiting the common bandwidth and at the same time shows a worse strain power spectral density; therefore, the $\sigma_{null}(\omega)$ defined in eq.\ref{cross-freq} is dominated by the strain noise of AURIGA unless $R_{AU}/R_V\gg 1$. This condition however is occurring rarely enough: $R_{AU}/R_V <2.4 $ with 90 \% probability for a galactic population of sources, and very similar values occur also considering the LIGOs or GEO interferometers in place of Virgo. Fig.\ref{xcorr-filter} shows the $\sigma_{null}(\omega)$, corresponding to $R_{AU}/R_V=\lbrace0,1,2.4\rbrace$ . These curves are quite similar and therefore it is advisable to approximate $R_\alpha = R_\beta$ in eq.\ref{cross-freq}, i.e. to consider filtered data streams:
\begin{equation}
\widetilde{x}_{\alpha, w}'(\omega) \equiv \frac{\widetilde{x}_{\alpha}}{\sqrt{S_{\alpha}(\omega) + S_{\beta}(\omega)}}
\end{equation}
to compute the cross-correlation. In addition, to be robust against noise fluctuations and relative calibration errors between AURIGA and Virgo, we considered the r-statistics of such filtered data streams: 
\begin{equation}
r_{\alpha, \beta} \equiv \frac{\int dt x_{\alpha, w}'(t) x_{\beta, w}'(t+t')}{\sqrt{\int d\tau x_{\alpha, w}'^2(t) \int d\tau x_{\beta, w}'^2(t)}}
\end{equation}
where we integrate over a time window of fixed duration and search for local maxima as a function of the time $t$ and time shift $t'$ between the data streams.

\section{Network Analysis}

We analysed the 24 hours of data starting from the UTC time 14 Sep 2005 23:11:27. 
In this time period, Virgo was performing its seventh commissioning run (C7) while AURIGA was in stable operation. This affects mainly the quality of the data of Virgo for the presence of periods of instrumental unlock
reducing the observation time from 24 to 16 hours and 33 mins. 
Moreover, the Virgo Collaboration provided a list of vetoes to flag the most noisy periods, corresponding to $14.5\%$ reduction of the live time.
The $S_{hh}$ sensitivity curves for the two detectors in the $[850\div950]$ Hz band on September $15^{th}$ are shown in fig.\ref{xcorr-filter}.

\subsection{Pipeline description}

\begin{figure}[!h]
 \begin{center}
  \includegraphics[width=150mm,height=50mm]{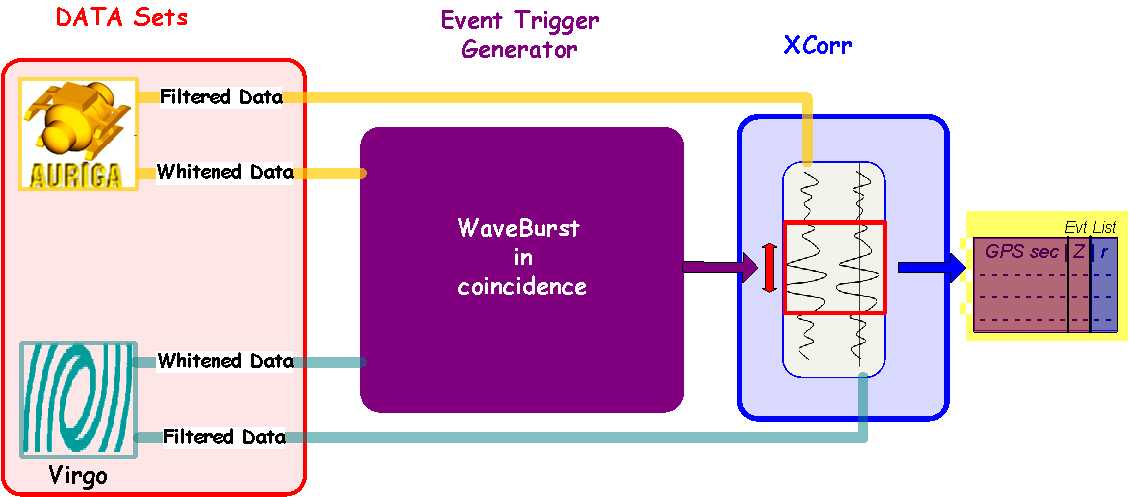}
 \end{center}
\caption{\small{Block diagram of the pipeline. The two  detectors whitened data sets are passed to WaveBurst that produces a list of coincident triggers, each characterized by the geometrical significance, $Z$. The XCorr takes the two weighted filtered data chuncks around the coincident trigger times and slides them looking for the maximum of the cross-correlation coefficient, $r$. }}\label{pipeline}
\end{figure}

Fig.\ref{pipeline} shows the scheme of the implemented pipeline:
as a first step, the raw data from the two detectors are whitened and bandpassed to match the AURIGA bandwidth  (i.e. 850 - 950 Hz), then they are fed in as input to WaveBurst \cite{Waveburst}. 

WaveBurst has been configured to search for coincident excesses of power on the two data sets over 4 wavelet decomposition levels, corresponding to time/frequency resolutions ranging from 4.916 ms by 101.72 Hz  to 39.32 ms by 12.72 Hz; it then produces the list of triggers, containing all the trigger parameters (e.g. the central time in GPS seconds, the central frequency, the frequency range, the geometric significance, $Z$\footnote{For the definition of the significance, see eq.s 10 and 11 in \cite{Waveburst}}, the Signal to Noise Ratio, SNR, etc. ); for each of those events, the last step of the pipeline, XCorr, selects the corresponding chuncks of weighted filtered data (see sec. \ref{method}) and, by sliding on $t$ and $t'$ the two time series \footnote{Between Virgo and AURIGA, the light travel time is $\leq 0.8 $ ms and the phase shift is $\leq 0.5 $ ms; to be conservative we set the relative time slide to $ |t'|<1.5$ ms.}, it searches the maximum of the cross-correlation coefficient, $ r $, computed over a given time window.

Within our approach, $Z$ and $r$ are the main test statistics which characterize each event at the output of the pipeline. A combined threshold on these quantities sets the overall false alarm rate (FAR) and detection efficiency of the network.  
%In the above sketched pipeline, there are a number of \textit{technical} parameters that have not been mentioned (i.e. WaveBurst and XCorr internal parameters), whose role is to optimize the pipeline. 
Other internal parameters have been set in order to have neglegible impact on the final FAR and detection efficiency.

\subsection{Background and detection efficiency estimation}
\label{background}
The background is estimated by applying the pipeline to 4000 time-shifted data sets with a total live-time of 2760 days, with shifts ranging from -2123s to +2123s in steps of $\approx 1$ s.
This choice allows to have a sufficiently large number of resamplings, while preserving the main characteristics of the data sets. 

The detection efficiency has been computed over a number of different waveforms (mainly elliptically polarized damped sinusoids) by means of software injections: 
the signal central frequency $f_0$ ranges within the bar bandwidth (850-950 Hz) and the decay time $\tau$ spans at most a few tens of milliseconds, with random inclination and polarization angles. We have considered source populations distributed at the Galactic Center, in our galaxy according to its mass distribution \cite{flynn1996kos} and uniform distributed in the sky.  

For each waveform, we have created two sets of 994 signals, approximately spaced by 60 s: the {\it tuning} and the {\it estimation} sets.

\subsection{Tuning of the analysis}\label{tuning}

The aim of the tuning phase is to optimize the pipeline parameters in order to achieve a low FAR, while preserving the detection efficiency.
The search pipeline has been tuned using 2000 time-shifts randomly selected out of the initial 4000 and the {\it tuning} set of injections.
The parameters set in this tuning phase have then been used on the second half of the time shifts to get the final FAR and on the {\it estimation} set of injections to calculate the network efficiency.%(fig.(\ref{EffGC-GD-AS})).

\paragraph{Cross-correlation Window}
Among the pipeline parameters, the cross-correlation window is one of the most important.
The general view is that a short time window may be unsuitable for a long signal, as a significant part of its power may be cut off by the window itself. At the same time, if the integration time span is too large, the signal may be diluted in it. Indeed, the weighted cross-correlation filter induces a correlation between the 2 data series over a time scale of the order $\simeq 30$ms. Therefore, even for delta-like signals, we need a cross-correlation window of at least $60$ ms. 
We have set it to 100 ms after testing the performances of various time windows ([25, 50, 100, 200, 400, 800]ms)
in terms of FAR and efficiency for narrow and wide band simulated signals.

\paragraph{Combined threshold}
We set the FAR to $\approx 2/yr$: given our short observation time, we could not target a smaller rate, since only 7 background events were surviving. We have not applied the Virgo list of vetoes, as from tests performed on the tuning set of the shifts, we measured a $\approx 30\%$ FAR reduction only, while losing $14.5\%$ of the live time. 

The target FAR has been achieved by setting a combined  ($Z, r$) threshold on the event list selected by the pipeline.
In fig.\ref{Eff 930-30, 914-14}, we show an example of how a population of injected signals (squares) detected by the pipeline is distributed with respect to the background events (dots) for two waveforms. The dashed line shows the optimized combined threshold on ($Z,r$): such threshold is defined by a simple linear equation: $Z=m r + q$, where the parameters $m, q$ are set in such a way as to achieve the maximum efficiency at our target FAR.
We underline here that, since AURIGA performance over the short time period of the selected data is quite stable, its noise distribution does not change significantly over different time spans and hence the choice of our threshold is fairly general. 
\begin{figure}[!h]
 \begin{center}
 \begin{tabular}{cc}
  \includegraphics[width=75mm, height=60mm]{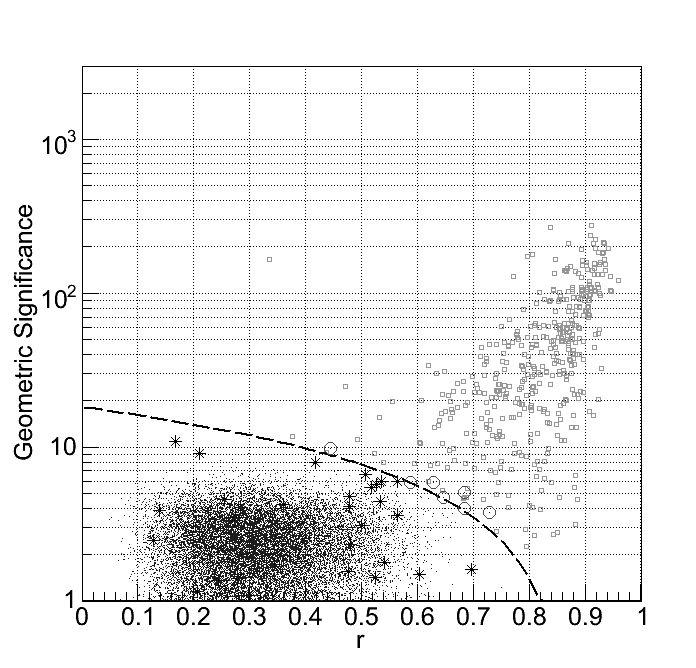} &
  \includegraphics[width=75mm, height=60mm]{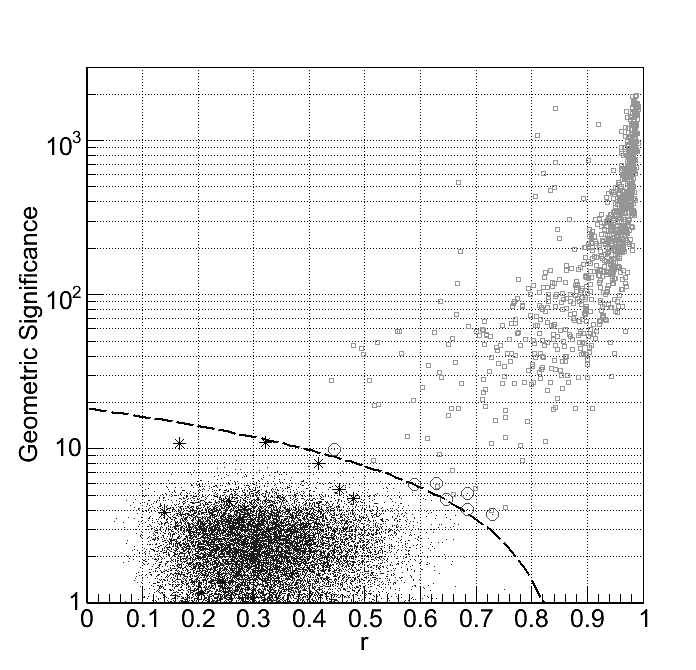}
  \end{tabular}
 \end{center}
\caption{\small{ Geometrical significance versus {\it r-statistic} for background events (dots) and two injected Damped Sinusoids (squares). a): DS $f=914$ Hz, $\tau=1$ ms, $h_{rss}=1\times 10^{-19}$ $1/\sqrt{Hz}$. b): DS $f=930$ Hz, $\tau=30$ ms, $h_{rss}=1\times 10^{-19}$ $1/\sqrt{Hz}$. Circled dots are background events resulting above threshold (dashed line) and hence contributing to our FAR. Asterisks are instead injected events not passing the threshold. }}
\label{Eff 930-30, 914-14}
\end{figure}

\section{Results}
\begin{figure}[!h]
  \begin{center}
\begin{tabular}{cc}
   \includegraphics[width=75mm]{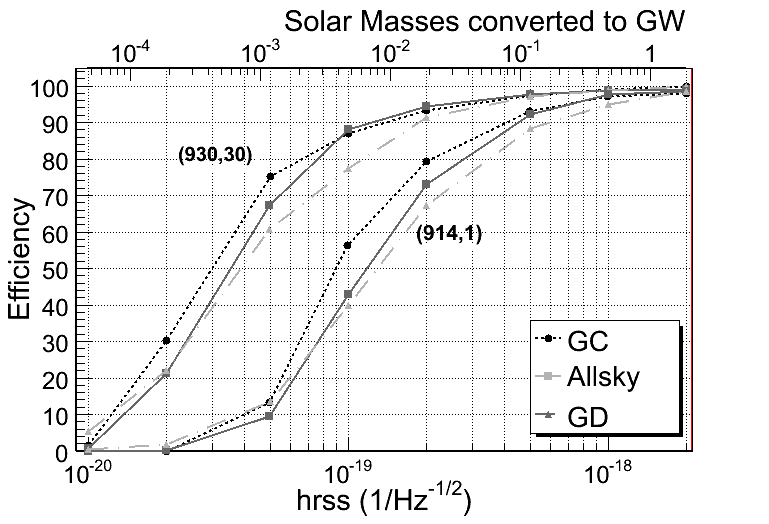}
\includegraphics[width=80mm]{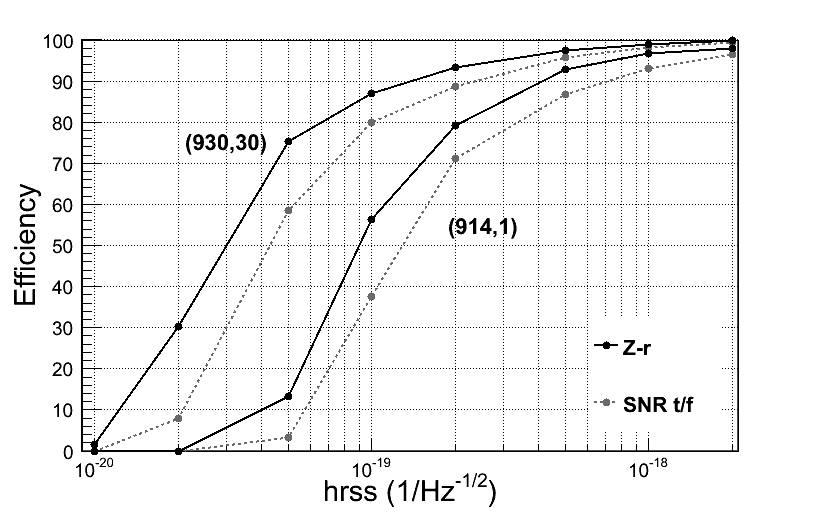}
\end{tabular}
\end{center}
\caption{\small{a): Detection efficiencies of the WaveBurst+XCorr pipeline vs $h_{rss}$ for Galactic Center, all sky and Galactic Disk source populations and for the two signals (930 Hz, 30ms and 914 Hz, 1ms). On the upper x-axis, the $h_{rss}$ is converted to energy emitted in GWs at Galactic Center, using eq.B6 in \cite{virgobars}. The network response is comparable for the 3 populations, but it is strongly dependent on the signal duration. b): Comparison between the detection efficiencies for the 2 pipelines at FAR $\simeq 2/yr$. Solid line: WaveBurst+XCorr pipeline with a combined threshold on ($Z,r$). Dashed line: time-frequency coincidence search with two separate thresholds on the SNR of the detectors. For the time-frequency search only, the Virgo vetoes were necessary to reduce the FAR, while the WaveBurst+XCorr search results to be robust enough to reduce the FAR also in presence of the extra noise associated to the vetoed time periods. The plotted efficiencies have been normalized to the corresponding live-times, accounting for the loss due to the vetoes. }}
\label{EffGC-GD-AS}
\end{figure}

By applying the \textit{tuned} analysis over the estimation set of the time-shifted resamples, we get a final $FAR=1.6/yr$ with a statistical sigma of $0.6/yr$. 
The detection efficiency of our network is shown in fig.\ref{EffGC-GD-AS}a, for the 3 injected source populations of damped sinusoids waveforms.
The $50\%$ network detection efficiency is achieved in the range $h_{rss}^{50\%} \in [~3\times 10^{-20}, ~1.3\times 10^{-19}] 1/\sqrt{Hz}$. % for the DS injected populations. 
The efficiency is slightly dependent on the source populations, while it strongly depends on the signal duration, as expected due to the AURIGA narrower bandwidth. 

\subsection{Comparison with the time-frequency coincidence pipeline}
\label{Comparison}
We have compared the results of this analysis (i.e. detection efficiency, FAR, observation time) with those of a simple time-frequency coincidence of single detector triggers.
In the latter pipeline, we have analysed the two data streams separately using WaveBurst as event trigger generator and the resulting triggers retained if coincident in time ($\Delta t=64 $ ms) and frequency ($\Delta f = 25 Hz$). We have then applied the list of vetoes for Virgo C7 data to the surviving triggers, reducing by a factor $\approx 5$ the accidental coincidences at the cost of $14.5\%$ of live time. Finally, an optimized threshold on SNR has been applied on the resulting events from the tuning set, in order to get the target FAR$\simeq 2/yr$. The best efficiency has been achieved by setting the threshold at $SNR > 5.6$ for AURIGA and $SNR > 14.5$ for Virgo. On the estimation sets, the FAR results $5.3 \pm 1.3/yr$ %$1.6\pm0.6/yr$ 
and the corresponding detection efficiencies are reported on in fig.\ref{EffGC-GD-AS}.b for a Galactic Center source population.
At similar FAR, the proposed pipeline outperforms the time-frequency coincidence for the tested waveforms in detection efficiency. 

\section{Conclusions}
We investigated the performances of a joint search for GW bursts by a narrow-band resonant detector and a wide-band interferometer using one day of data taken by AURIGA and Virgo in September 2005. The data analysis method based on cross-correlation outperforms a simpler time-frequency coincidence search: it achieves better efficiencies at equal false alarm rate, mostly because it is by far more powerful in discriminating accidental coincidences.

This search monitors at 50\% efficiency galactic sources emitting $\approx7\times 10^{-4} \div 7\times 10^{-3} M_\odot c^2$  in GW bursts of $30\div 1$ ms decay time, provided that their strongest Fourier components are in the AURIGA bandwidth. 
The resulting detection efficiency of the hybrid network is limited by the less sensitive detector, AURIGA, to $5\div 10$ times larger amplitudes with respect to a Virgo only search on a larger C7 data set \cite{C7allsky} for pulses of $\sim 1ms$ duration. On the other hand, there is a clear advantage of this hybrid search in the background reduction with respect to a single detector search, as it allows to identify candidate events with high statistical confidence.

The main limitations of this methodological study comes from the short duration of the data set used, which prevented us from investigating false alarm rates lower than a few per year. We expect that the efficiency of this methodology would take only a small benefit from the achieved progresses in sensitivity of the interferometric detectors Virgo, LIGOs and GEO; however, these progresses should reduce significantly the false alarm rate of this search. Further studies with newer and longer data sets are necessary to assess quantitatively this issue.  
Nevertheless, this hybrid network search could be of interest for the near future, when only one interferometer will be taking data, a likely condition for a large fraction of the time due to the planned instrumental upgrades towards enhanced detectors.

%%%%%%%%%%%%%%%%%%%%%%%%%%%%%%%%%%%%%%%%%%%%%%%%%%%%%%%%%%%%
%\begin{center}
 \section*{References}
%\end{center}

%%%%%%%%%%%%%%%%%%%%%%%%%%%%%%%%%%%%%%%%%%%%%%%%%%%%%%%%%%%%
%\section*{References}
%%%%%%%%%%%%%%%%%%%%%%%%%%%%%%%%%%%%%%%%%%%%%%%%%%%%%%%%%%%%

%\bibliographystyle{unsrt}
%\bibliographystyle{iopart-num}
%\newpage %Just because of unusual number of tables stacked at end
%\bibliography{AURIGAVirgo_draft_CQG}% Produces the bibliography via BibTeX.

%%%%%%%%%%%%%%%%%%%%%%%%%%%%%%%%%%%%%%%%%%%%%%%%%%%%%%%%%%%%
\providecommand{\newblock}{}

\end{document}